\documentclass[preprint,showpacs,amsmath,amssymb,floatfix,pre]{revtex4}
\usepackage{graphicx}
\begin{document}
\title{The spin-3/2 Blume-Capel model with competing short- and long-range interactions} 

\author{Octavio D. Rodriguez Salmon \footnote{Corresponding author.\\ E-mail address: octaviors@gmail.com}}

\affiliation{Universidade Federal do Amazonas, Departamento de F\'{\i}sica, 3000, Japiim, 69077-000, Manaus-AM, Brazil}

\author{Jos\'{e} Ricardo de Sousa}
\affiliation{Departamento de F\'{\i}sica, Universidade Federal do Amazonas, 3000, Japiim,
69077-000, Manaus-AM, Brazil}
\affiliation{National Institute of Science and Technology for Complex Systems, 3000, Japiim,
69077-000, Manaus-AM, Brazil}

\author{Minos A. Neto } 
\affiliation{Universidade Federal do Amazonas, Departamento de F\'{\i}sica, 3000, Japiim, 69077-000, Manaus-AM, Brazil}

\author{Igor T. Padilha } 
\affiliation{Universidade Federal do Amazonas, Departamento de F\'{\i}sica, 3000, Japiim, 69077-000, Manaus-AM, Brazil}

\author{J. Roberto Viana Azevedo} 
\affiliation{Universidade Federal do Amazonas, Departamento de F\'{\i}sica, 3000, Japiim, 69077-000, Manaus-AM, Brazil}

\author{Francisco Din\'{o}la Neto } 
\affiliation{ University of Cambridge; Cavendish Laboratory, J. J. Thomson Avenue, Cambridge, CB3 0HE, England.}

\date{\today}

\begin{abstract}
\noindent
The phase diagrams of the spin-$3/2$ Blume-Capel model with competing short and long-range interactions were studied through the free energy density obtained by analytical methods. The competition emerges when positive short-range interactions of strength $K$ arranged in a linear chain tend to establish an anti-parallel spin order, whereas negative long-range interactions $-J$ tend to align them in parallel. Thus, no  ferromagnetic order exists for $K/J > 0.25$. So, the phase-diagrams were scanned by varying the values of $K$ in this interval.  As  in other similar study done for the spin-1 case, the second-order frontier separating the ferromagnetic and the paramagnetic phases is transformed gradually into a first-order line,   when    $K/J$ is greater than a  certain critical value. Accordingly, there is a subinterval of $K$, for which  two tricritical points appear restricting the length of the second-order frontier. Nevertheless, for greater values of   $K/J$,  the ferromagnetic-paramagnetic frontier becomes wholly of first order. Also, the tipical coexistence line, which divides two different ferromagnetic phases of magnetization $m=3/2$ and $m=1/2$, becomes more complex by giving rise  to another  line of coexistence with a reentrant behavior that   encloses a third ordered phase. In this case, the competition is such that there is a region in the phase diagram, where  for each spin $i$ with $S_{i}=3/2$ ($S_{i}=-3/2$), there is another one spin $j$ with $S_{j}=-1/2$ ($S_{j}=1/2$), so the absolute value of the magnetization per spin is one. 
\vskip \baselineskip
%\vspace{2cm}
%\medskip

\noindent
Keywords: Spin-3/2, Ising Model, Multicritical Phenomena, Blume-Capel Model.
\pacs{05.70.Fh, 05.70.Jk, 64.60.-i, 64.60.Kw}

\end{abstract}
\maketitle
\newpage

\section{Introduction}

In solid structures, the competition arises when two or more physical parameters tend to favor states with different symmetry, periodicity or structure. Accordingly, this kind of competition creates interesting magnetic phases. For instance, the crystal cerium antimonide $\bf CeSb$ is a $\bf NaCl$-like alloy, in which  ions of $\bf Ce$ and $\bf Sb$ occupy alternate vertices of  a cubic lattice. Its phase-diagram shows various magnetics structures \cite{mignod} that have been explained by the ANNNI model, whose ferromagnetic nearest-neighbor interaction favours a homogeneous arrangement of spins, while the antiferromagnetic coupling prefers the periodic arrangement of two spins up, two down and so on \cite{FISHER,Cristian}. Thus, systems with competing interactions show interesting properties
\cite{diep, si,lacroix,fischer,jr1,salmon1}.  
\vskip \baselineskip
\noindent

 From the theoretical point of view, there is  a special interest in studying spin models with competing short- and long-range interactions. It is important to mention that  the Ising model in the mean-field approach is equivalent to the Ising model in which all pairs of spins are  coupled with the same  constant $J/N$, where $N$ is the total number of spins \cite{thomson}. Baker  reported that Siegert was the first to show him this fact \cite{baker}.  Therefore, these are called mean-field interactions or infinite-range interactions. Their utility is for representing long-range interactions due to the fact that the Ising model is exactly solved with them.  Also, Baker   showed that even in the presence of short-range interactions, the existence of any coupling  of infinitely long range is sufficient to change the  nature of the transition to be that of  mean-field type \cite{baker}. 
\vskip \baselineskip
\noindent

 There is also an  interest in investigating spin models with  competing ferromagnetic and antiferromagnetic couplings of long and short range,  motivated by the multicritical behavior that may appear.   An early attention for this kind of competition was given by Nagle \cite{nagle} who showed the existence of a  spontaneous magnetization between two non-zero temperatures  in a linear spin$-1/2$ chain. So, if the ferromagnetic couplings have longer range than the antiferromagnetic ones, the ferromagnetic interactions are strong enough to induce order at some temperature interval. Another similar work for the Ising model  is found in a paper published by Kardar \cite{kardar}. There he studied a competition between mean-field ferromagnetic interactions  and nearest-neighbor interactions for dimension $d=1,2$. If the nearest-neighbor interactions are antiferromagnetic, there is a frontier line in the phase diagram  with a tricritical point separating the ferromagnetic phase and the disordered phase, for $d=1$. For $d=2$, this frontier becomes richer, because now it separates the ferromagnetic phase and two phases with zero magnetization, namely, the antiferromagnetic phase and the disordered phase. This model was called the Nagle-Kardar Model  \cite{ bonner,kaufman, mukamel,cohen}. Also,  Hamiltonians like this,  with competing local, nearest-neighbor, and mean-field couplings, have also been solved in both, the canonical and the microcanonical ensemble so as to test ensemble inequivalence \cite{campa,duv}. 
\vskip \baselineskip
\noindent

In what Ising spin-1 models concerns, it is important to quote the work of  U. Low, et al. \cite{low}, who did a coarse-grained representation of frustrated phase separation in high temperature superconductors, by using the following Hamiltonian:

\begin{equation}
H = \frac{Q}{2}\sum_{i \neq j}\frac{S_{i}S_{j}}{r_{ij}} - L\sum_{<ij>}S_{i}S_{j} + K \sum_{j}S_{j}^{2}
\end{equation}

where $S_{j}=0,\pm 1$ are spins in a square lattice, and $L,Q > 0$. Note that the spins in the first term are coupled by positive long-range interactions of Coulomb type, whereas in the second term the spins are coupled with negavite  nearest neighbors interactions. So, it emerges a competition between the two terms,  which  tend to align the spins in ferromagnetic  and antiferromagnetic order. The third term is the term of anisotropy, which controls the number of sites at which $S_{j}^2 = 1$.  The authors found that the ground state presents a complex phase diagram with a rich  variety of phases in the $Q/K - L/K$ plane, for $K>0$.  Recently, a similar Hamiltonian was studied for finite temperatures  by Salmon, Sousa and Neto \cite{salmon2}, though they considered mean-field couplings for the ferromagnetic interactions, and one-dimensional nearest-neighbor couplings for the antiferromagnetic interactions. In this case, the expression of the free energy was obtained by using analytical methods.
\vskip \baselineskip
\noindent

In this work we improve the research of this  type of competition for the spin-$3/2$ Blume-Capel Model \cite{blume,capel}. In the following section we  present the Hamiltonian and the details of its Statistical Mechanic treatment.

\section {The Hamiltonian and the free energy}

We consider a spin-$3/2$ chain with long-range and short-range competing interactions represented by the following Hamiltonian:

\begin{equation}
\label{hamiltonian}
{\cal H} = -\frac{J}{2N} ( \sum_{i=1}^{N} S_{i}  )^{2} + K \sum_{i=1}^{N}S_{i}S_{i+1} + D \sum_{i=1}^{N}S_{i}^{2},
\end{equation}
 \vskip \baselineskip
\noindent
where $S_{i} = -3/2,-1/2,1/2,3/2$, for  $i=1,2...,N-1,N$, being $N$  the number of spins. The first sum represents the mean-field ferromagnetic interactions, thus,  each spin $S_{i}$ interacts equally with all the $N$ spins (itself included), by  couplings of strength $J$. This first sum is responsible for the ferromagnetic order because we set $J>0$.   The second sum represents the energy of a linear chain of spins interacting between their nearest-neighbors with  coupling constant $K$. In order  to create a competition between the short-range antiferromagnetic interactions and the  long-range ferromagnetic couplings of the first sum, we consider $K>0$. The last sum is the anisotropy  term with constant $D$ ($D>0$).  For $K=0$, we recover the spin-$3/2$ Blume-Capel  Model with mean-field ferromagnetic interactions, which is a particular case of the Blume-Emery-Griffiths (BEG) model, where $S=3/2$ \cite{beg1,beg2}. It is important to mention that the BEG model, for $S=3/2$, with dipolar and quadrupolar  interactions was introduced to explain phase transitions in the $ \bf DyVO_{4}$ compound \cite{cooke}. Also,  the  Blume-Capel model for $S=3/2$ has attracted the antention for its multicritical behavior when considering $D/J$ as a random variable \cite{bahmad} and when implemented in a two-dimensional lattice with antiferromagnetic interactions in the presence of an external magnetic field \cite{smaine}. 
\vskip \baselineskip
\noindent
As a previous  step to  obtain  the phase diagrams of  this new version of the Nagel-Kardar model, we have to determine the analytical expression of the free energy. To this end, we firstly need to calculate the partition function $Z$ in the canonical ensemble \cite{huang}:

\begin{equation}
Z=   {\rm tr} \left \{ e^{- \beta {\cal H}} \right \}~, 
\end{equation}
\vskip \baselineskip
\noindent
where $\beta = \frac{1}{k_{B}T}$, $k_{B}$ is the Boltzman constant,  $T$ stands for the temperature of the system, and $\displaystyle {\rm tr \{...\}} \equiv \sum_{S_{1}=-1}^{1} \sum_{S_{2}=-1}^{1}...\sum_{S_{N}=-1}^{1} $  indicates the sum over all spin configurations. In this class of interaction the Hubbbard-Stratonovich   transformation ~\cite{hubbard, dotsenkobook} can be applicable  to decouple the spins in the quadratic term in Eq.~(\ref{hamiltonian}). Accordingly, this transforms the partition function as follows

\begin{equation}
Z =  \sqrt {\frac{N \beta J}{2\pi}} \int_{-\infty}^{\infty} \left  \{ e^{(-\frac{1}{2}\beta N J x^{2})} \sum_{\{ S_{i} \}} \prod_{i=1}^{N}  e^{ \beta  Q(i,i+1)} \right \}  dx  ~,
\end{equation}
\vskip \baselineskip
\noindent
where  $Q(i,i+1) =  \frac{1}{2}J x (S_{i}+S_{i+1}) - K S_{i}S_{i+1} -  \frac{1}{2} D (S_{i}^{2}+S_{i+1}^{2})$. So, the partition function can be now calculated by using the transfer matrix technique:

 \begin{equation}
Z =  \sqrt {\frac{N \beta J}{2\pi}} \int_{-\infty}^{\infty} \left \{ e^{-\frac{1}{2}\beta N J x^{2}} {\rm Tr} \{ {\bf M}^{N} \} \right \}  dx  ~,
\end{equation}
\vskip \baselineskip
\noindent
where $ \bf M$ is the matrix transfer, given by:
\begin{equation} 
\label{transferM}
{\bf M} = \begin{bmatrix}
 e^{-\beta(\frac{3}{2}Jx+\frac{9}{4}K+\frac{9}{4} D)}  & e^{-\beta(Jx+\frac{3}{4}K +\frac{5}{4}D)} & e^{\beta(-\frac{1}{2}Jx + \frac{3}{4} K-\frac{5}{4} D)} & e^{\beta(\frac{9}{4}K-\frac{9}{4}D)} \\
 e^{-\beta(Jx+\frac{3}{4}K +\frac{5}{4}D)}      & e^{-\beta(\frac{1}{2}Jx+\frac{1}{4}K +\frac{1}{4}D)} & e^{\beta( \frac{1}{4} K-\frac{1}{4} D)} & e^{\beta(\frac{1}{2}Jx + \frac{3}{4}K-\frac{5}{4}D)} \\
 e^{\beta(-\frac{1}{2}Jx + \frac{3}{4} K-\frac{5}{4} D)}  &  e^{\beta( \frac{1}{4} K-\frac{1}{4} D)} & e^{\beta(\frac{1}{2}Jx - \frac{1}{4}K-\frac{1}{4}D)} & e^{\beta(Jx - \frac{3}{4}K-\frac{5}{4}D)} \\
 e^{\beta(\frac{9}{4}K-\frac{9}{4}D)}  & e^{\beta(\frac{1}{2}Jx + \frac{3}{4}K-\frac{5}{4}D)} & e^{\beta(Jx - \frac{3}{4}K-\frac{5}{4}D)} & e^{\beta(\frac{3}{2}Jx - \frac{9}{4}K-\frac{9}{4}D)} \\
\end{bmatrix}
\end{equation}
\vskip \baselineskip
\noindent
The trace ${\rm Tr} \{ {\bf M}^{N} \} $ is equal to ${\displaystyle \sum_{j=1}^{4}} \lambda_{j}^{N}$, where $ \{ \lambda_{j} \}$ are the  eigenvalues of $\bf M$. In  the thermodynamic limit ($N \to \infty $), the partition function is simplified through the steepest descent method, so:

\begin{equation}
Z =  \int_{-\infty}^{\infty} \left \{ \displaystyle e^{ -N(\frac{1}{2}\beta J x^{2} -\frac{1}{N} \log ( \sum_{j=1}^{4}\lambda_{j}^{N}) -\frac{1}{2N}\log(\frac{N J \beta}{2 \pi}))}  \right \}  dx \simeq  e^{ -N\beta f } ,
\end{equation}
\vskip \baselineskip
\noindent
where 

\begin{equation}
\label{fdensity}
f =   \frac{1}{2}  J m^{2} - \frac{1}{\beta} \log(\lambda_{max}), 
\end{equation}

being $\lambda_{max}$ equal to  $\max \{ \lambda_{1},\lambda_{2},\lambda_{3},\lambda_{4}  \} $, and $m$ is the value of $x$ that minimizes the function $f$, called the free energy density, for given values of $k_{B}T/J$, $D/J$ and $K/J$. It can also  be proven that $m$ is the magnetization of the system at the equilibrium (see the Appendix of reference \cite{salmon2}). 
\vskip \baselineskip
\noindent

Now, with the aid of the free energy density $f$, we can explore the ferromagnetic frontiers and their limits in the  $k_{B}T/J-D/J$ plane, knowing that the antiferromagnetic phases are only present for $T=0$. So, the only relevant order parameter is  the magnetization, which was calculated  by finding the minima of the function $f$ as a function of $m$  in  Eq.(\ref{fdensity}), for given values of $k_{B}T/J$, $D/J$ and $K/J$.  We  obtained, numerically, the maximum eigenvalue of  the transfer matrix of Eq.(\ref{transferM}).  In order to estimate the points $(D/J,k_{B}T/J)$ belonging to the  ferromagnetic-paramagnetic frontiers and those ones  that divide different ferromagnetic phases, we scanned the magnetization curve $m$ versus $k_{B}T/J$, by fixing $K/J$ and $D/J$, and the curve $m$ versus $D/J$, for fixed values of $K/J$ and $k_{B}T/J$. The type of phase transition was determined by analyzing the behavior of the magnetization and free energy at the frontier points. First-order points are those at which the magnetization suffers a discontinuous change due to the coexistence of different phases, whereas at second-order points the magnetization is   continuous.

To plot the frontiers and points of  the phase diagrams we  use distinct symbols, as described below (see Reference \cite{griffiths}).   

\begin{itemize}

\item Continuous (second order) critical frontier: continuous line;

\item First-order  frontier (line of coexistent): dotted line;

\item Tricritical point: located by a black circle;

\item Ordered critical point: located by a black asterisk;

\end{itemize}

\section{Reviewing the  $K=0$ case}

The typical spin-$3/2$ Blume-Capel model with mean-field ferromagnetic is recovered by setting $K=0$, in the Hamiltonian given in Eq.(\ref{hamiltonian}). For this case, the explicit expression of the free energy density in Eq.(\ref{fdensity}) can be easily  written down as

\begin{equation}
f=\frac{1}{2}Jm^{2} - \frac{1}{\beta}\log \left [ 2e^{-\frac{9}{4}\beta D} \cosh(\frac{3}{2}\beta Jm) + 2e^{-\frac{1}{4}\beta D}\cosh(\frac{1}{2}\beta Jm) \right ].
\end{equation}
 \vskip \baselineskip
\noindent
The   Landau expansion of the above free energy density is:

\begin{equation}
f = f_{0} + a m^{2} + b m^{4}+....,
\end{equation} 

where 

\begin{equation}
f_{0} = - \frac{1}{\beta} \log(2e^{-\frac{9}{4}\beta D} + 2e^{-\frac{1}{4}\beta D}),
\end{equation}

\begin{equation}
a=J-\frac{1}{4}\beta J^{2} (\frac{9+e^{2\beta D}}{1+e^{2\beta D}}),
\label{coefa}
\end{equation}
 \vskip \baselineskip
\noindent
The explicit expression of the coefficient $b$ is too lengthly to be written here. The magnetization $m$ is obtained by extremizing the function $f$,  $\partial f/\partial m = 0$, that leads to the following transcendental equation:

\begin{equation}
F(m) = 0, 
\end{equation}

where 
\begin{equation}
F(m) = m - \frac{3 \sinh(\frac{3\beta Jm}{2}) + e^{2\beta D} \sinh(\frac{\beta Jm}{2})}{2 \cosh(\frac{3\beta Jm}{2}) + 2 e^{2\beta D} \cosh(\frac{\beta Jm}{2})} 
\end{equation}
 \vskip \baselineskip
\noindent
The second-order frontier of the phase diagram is plotted after solving numerically the equation $a=0$, with the condition $b>0$. There is also a first-order line separating two ferromagnetic phases $\bf F_{1}$ and $\bf F_{2}$, with diferent values of the magnetization per spin ($m=m_{1}$ and $m=m_{2}$). Thus, this frontier is obtained by solving the following non-linear set of equations, by the Newton-Raphson method:

\begin{equation}
f(m= m_{1}) = f(m = m_{2}),
\end{equation}

\begin{equation}
F(m=m_{1}) = 0,
\end{equation}

\begin{equation}
F(m=m_{2}) = 0,
\end{equation}

  \vskip \baselineskip
\noindent
with the initial conditions $T=0.005$, $D=0.5$, $m1=1.5$ and $m_{2}=0.5$,  setting $J=1$ and $k_{B} =1$. Accordingly, the corresponding phase diagram in the $D/J-k_{B}T/J$ plane is shown in Figure 1 (see also Fig. 2 in reference \cite{barreto}). There we can see a second-order frontier whose critical points separate the ferromagnetic phases $\bf F_{1}$  and $\bf F_{2}$, and the paramagnetic phase $\bf P$ ($m=0$). At low temperatures, the order parameter $m$, takes the values $m=3/2$ and $m=1/2$, for phases $\bf F_{1}$ and $\bf F_{2}$, respectively.   When $D/J \to \infty$,  the critical temperature of this frontier remains  constant, being  $T=0.25 J/k_{B}$. This can be shown by solving the equation $a=0$ (see Eq.(\ref{coefa})), when $D/J \to \infty$. The two ordered phases $\bf F_{1}$ and $\bf F_{2}$ are divided by a first-order frontier represented by a dotted line. It begins at $(D/J,k_{B}T/J)=(0.5,0.0)$ , and finishes at and ordered critical point located approximately at $(D/J,k_{B}T/J)=(0.4880(2),0.2890(2))$, represented by the asterisk.    Although this phase diagram has already been shown in past works \cite{beg2,barreto}, we noted an interesting behavior of the magnetization curve, after crossing through the first-order line. To illustrate this singular behavior,  we show in Figure 2a a short interval of $D/J$, where we can visualize better the zone of the phase diagram where this line of coexistence appears. The arrow is a guide to the eyes to show the vertical line at wich the magnetization was plotted in Figure 2b. In this case the arrow begins at $D/J = 0.494$. Accordingly, in Figure 2b is shown the magnetization versus the temperature, for the convenient value $D/J=0.494$. The jump discontinuity shows that the magnetization curve has crossed the line of coexistence.   Interestingly, we can observe  that the magnetization curve increases slightly with the temperature, after crossing this frontier, until reaching a maximum value. This happens only for values of $D/J$ in the interval for which the line of coexistence is present ($D/J \simeq 0.5$). Then the magnetization curve decreases continuously until falling downto zero, signaling that it has crossed the second-order critical frontier separating phases $\bf F_{2}$ and $\bf P$ (see where the arrow crosses the continuous line in Figure 2a). 
\vskip \baselineskip
\noindent
This review  of  the spin-$3/2$ Blume Capel with mean-field ferromagnetic couplings is  useful to understand how the topology of the phase diagram will evolve  by the addition of the second term of the Hamiltonian given in Eq.(\ref{hamiltonian}). So, as a previous step, we  obtain the phase diagram for $T=0$ and $K \geq 0$, in the next section. 

\section{The Ground State for $K \geq 0$}

At zero temperature,  the free energy is simply the energy ${E}$ corresponding to the Hamiltonian given in Eq.(\ref{hamiltonian}). Thus,  we have  to determine the spin configurations that minimize  this energy so as  to  obtain the phase diagram  in the  $D/J-K/J$ plane, for $D>0$ and $K>0$. It is easy to realize that there are four magnetic phases which give us four different values of ${E}$, that we denote as ${ E}_{\bf F_{1}}$, ${ E}_{\bf AF_{1}}$, ${ E}_{\bf F_{2}}$ and ${ E}_{\bf AF_{2}}$. Phases  $\bf F_{1}$ and $\bf AF_{1}$ denote the ferromagnetic and antiferromagnetic orders for which  the   spins have the absolute value $|S_{i}|=3/2$, for $i=1,2,...,N$. On the other hand, phases  $\bf F_{2}$ and  $\bf AF_{2}$ denote the ferromagnetic and antiferromagnetic orders for which  the spin variables have the absolute value  $|S_{i}|=1/2$, for $i=1,2,...,N$. Futhermore, the corresponding expressions of $E$ depend on the parameters $J$, $D$ and $K$, and are obtained according to the Hamiltonian presented in Eq.(\ref{hamiltonian}). Therefore, the  respective energy densities are the following:

\begin{equation}
\label{EF1}
{ E}_{\bf F_{1}}/N = -\frac{9}{8}J+\frac{9}{4}K+\frac{9}{4}D,
\end{equation}

\begin{equation}
\label{EAF1}
{ E}_{\bf AF_{1}}/N = -\frac{9}{4}K+\frac{9}{4}D,
\end{equation}

\begin{equation}
\label{EF2}
{ E}_{\bf F_{2}}/N = -\frac{1}{8}J+\frac{1}{4}K+\frac{1}{4}D,
\end{equation}

\begin{equation}
\label{EAF2}
{ E}_{\bf AF_{2}}/N = -\frac{1}{4}K+\frac{1}{4}D.
\end{equation}

Now we can  get the first-order frontiers separating these different phases by using the above expressions. Thus, the frontier dividing phases $\bf F_{1}$ and $\bf F_{2}$ is obtained by equating the expressions of Eq.(\ref{EF1}) and Eq.(\ref{EF2}), resulting in a line whose equation is given by $D/J = 1/2-K$. Similarly, we get the frontier dividing $\bf AF_{1}$ and $\bf AF_{2}$, which is described by the linear equation $D/J = K/J$, after equating the expressions  Eq.(\ref{EAF1}) and Eq.(\ref{EAF2}). We also found a vertical line, where $K/J = 1/4$, that separates phases $\bf F_{1}$ and $\bf AF_{1}$, as well as phases $\bf F_{2}$ and $\bf AF_{2}$, by equating Eq.(\ref{EF1}) and Eq.(\ref{EAF1}), as well as Eq.(\ref{EF2}) and Eq.(\ref{EAF2}). In Figure 3 we show these three frontiers meeting themselves at  a  point of coexistence represented by an empty triangle. 
\vskip \baselineskip
\noindent
In the next section we describe the results at finite temperatures, for $K>0$. It is important to mention that  phases $\bf AF_{1}$ and $\bf AF_{2}$ disappear when $T>0$. This is because both are caused by the nearest-neighbor antiferromagnetic couplings $K$ in the linear chain, then these long-range orders are destroyed in $d=1$, for $T>0$.  

\section{Results for $K>0$ and $T>0$ }
The frontiers of the phase diagrams for finite temperatures, are obtained by scanning the magnetization per spin $m$ throughout the $D/J-k_{B}T/J$ plane, for given values of  $K$ ($K>0$). The magnetization per spin, is the only relevant order parameter, which the free energy density $f$ depends on. So, for given values of $D/J$, $k_{B}T/J$ and $K/J$, we estimate numerically the value(s) of $m$, for which $f$ has its global minimum (or minima). In this way we can get numerically  vertical and horizontal curves of  $m$ in the $D/J-k_{B}T/J$ plane, so as to determine diferent types of frontiers, namely, first-order and second-order lines, for a given value of $K/J$. Due to the  progress of  computational  power in current machines, we noted that this is an efficient way to treat directly with the free energy density, for obtaining the phase diagrams. 
\vskip \baselineskip
\noindent
In what follows we present how the phase diagram of the spin-$3/2$ Blume Capel model with mean-field ferromagnetic interactions evolves when the antiferromagnetic coupling $K>0$ is taken into account (see the second term in the Hamiltonian given in Eq.(\ref{hamiltonian}). Firstly, we show in Figure 4 how the second-order frontier which separates the orderes phases and the paramagnetic phase (see Figure 1) suffers when  $K$ is increased. For lower values of $K/J$ (as for $K/J=0.15$), the frontier remains of second-order, but for greater values, such as $K/J=0.18$, the frontier is divided into three sections. The second-order section is limited by two tricritical points, and the sections of the extremes are of first order. We estimated that for $K/J=0.1758 \pm 0.0002$, the tricritical points begin to appear, and they approach themselves as $K/J$ increases, reducing the length of the second-order section. Then, for $K/J=0.22495 \pm 0.00005$, the tricritical points meet themselves, and for  $K/J>0.225$, the frontier separating the paramagnetic and ferromagnetic phases is only of first-order, as Figure 4 shows. 
\vskip \baselineskip
\noindent
The increase of $K/J$ affects also the coexistence line that divides the ordered phases  $\bf F_{1}$ and $\bf F_{2}$. For example, in Figure 5 we show the phase diagram for $K/J=0.22$. In Figure 5a we  see that another line of coexistence (ending at an ordered critical point) has emerged, like a branch,   from the line divding phases  $\bf F_{1}$ and $\bf F_{2}$ (see the region enclosed by the circle). In a similar work for the spin-$3/2$ Blume Capel model, branches like this has been reported (see the Fig.3 in reference \cite{baran}).  In Figure 5b we visualize more clearly the portion of the phase diagram containing this branch line. It begins at a point of coexistence represented by an empty diamond, and encloses a region of a third ordered phase, which we denote as $\bf F_{3}$.  Accordingly, this branch line divides phases $\bf F_{2}$ and $\bf F_{3}$ until its ending point. Its onset has been detected by scanning the magnetization curve versus temperature, for different values of $D/J$, for given values of $K/J$. For instance, in Figure 5b, the range in which this line is included is of width $\Delta D/J \simeq 0.0015$. However, for $K/J=0.190$, this width is of course shorter, as can be deduced from Figure 6, where the magnetization curve has been plotted for three close values of $D/J$. There we observe that for $D/J=0.30430$ and $D/J = 0.30440$, the magnetization curve is continuous, but for $D/J = 0.30435$, this suffers three jump discontinuities, which is a signal of the presence of the branch line. So, for $K/J=0.190$, the width of the range of the branch line is $\Delta D/J < 0.0001$. So the onset of the branch line must be for $K/J=0.190$, but very close to this value. Thus, for $K/J = 0.185$, we proceeded similarly for seeking the intervale of its appearance in the phase diagram, but the branch  line was not found in it. Therefore, its onset is estimated for $K/J = 0.1875 \pm 0.0025$. 
\vskip \baselineskip
\noindent
The region of the phase diagram containing the richest portion of this topology is especially analyzed in Figure 7, for $K/J=0.208$. So, in Figure 7a we may note the reentrant behavior of the branch line.  The arrows are guides to the eye for signaling where the mangentization is plotted in Figures 7c and 7d. Figure 7b is intended to show, through the free energy density, that at the point  represented by the empty diamond, whose coordinates are $(D/J,k_{B}T/J) = (0.2872040(2), 0.104310(2))$,  phases  $\bf F_{1}$, $\bf F_{2}$ and $\bf F_{3}$ coexist. This is why the free energy density is equally minimized by six values of $m$. For phase $\bf F_{1}$, $m$ is close to $1.5$, for phase $\bf F_{2}$, $m$ is close to $0.5$, and for  phase $\bf F_{3}$, $m$ is equal to 1.   On the other hand, the magnetization curves in Figures 7c and 7d show the first-order nature of points belonging to the coexistence lines. In Figure 7c, the magnetization as a function of the temperature, is plotted for $D/J=0.2874$ (see the vertical arrow in Figure 7a). It suffers three jump discontinuities , because it has crossed the first-order line separating phases $\bf F_{1}$ and $\bf F_{2}$, and the  reentrant zone of the branch line dividing phases $\bf F_{2}$ and $\bf F_{3}$. This is why there is a short magnetization gap between phases $\bf F_{1}$ and $\bf F_{3}$, where phase $\bf F_{2}$ is present. For greater values of the temperature, the magnetization falls continuously downto zero  due to the presence of the second-order section of the frontier dividing phases $\bf F_{2}$ and $\bf P$ (not shown in Figure 7a). In Figure 7d, we plotted $m$  versus $D/J$, for  $k_{B}T/J = 0.1173$, so as to study the behavior of $m$ along the horizontal line marked by the horizontal arrow shown in Figure 7a. This line crosses the vertex of the reentrant curve of the branch frontier. Thus, the magnetization suffers a jump dicontinuity  when passing through the transition from phase $\bf F_{3}$ to phase $\bf F_{2}$. 
\vskip \baselineskip
\noindent
For greater values of $K/J$, the reentrant form of the branch line disappears. Also,  the ending points of the two lower  frontiers and the upper  frontier approach themselves, as $K/J$ increases. This can be observed in Figure 8, where we  see the asterisks very close to the first-order frontier that separates the ferromagnetic and paramagnetic phases. Consequently, if  $K/J$ is greater than certain critical value $K^{*}/J=0.24585 \pm 0.00005$, these ending points get to touch the upper first-order frontier. In Figure 9 we show this fact in the phase diagram obtained for $K/J=0.248$. There the ending points of the lower first-order frontiers are now points of coexistence, and these are represented by an empty square and a black square.  Thus, the three first-order frontiers completely enclose the phase $\bf F_{3}$. So, for $K/J=0.248$, the coordinates of the point represented by empty square are $(D/J,k_{B}T/J) = (0.2430(2),0.069850(2))$, and  for the black square these are $(D/J,k_{B}T/J) = (0.249740(2),0.055910(2))$. In order to show the critical behavior at this points of coexistence, we plotted the free energy density at each of them. So, in Figure 10a, we observe five values of the magnetization  equally minimizing the free energy at the point represented by the empty square in Figure 9, showing that phases  $\bf F_{1}$, $\bf F_{3}$ and $\bf P$ coexist. Similarly, the free energy density plotted in Figure 10b is intended to show  that phases $\bf F_{3}$, $\bf F_{2}$ and $\bf P$ coexist at the point represented by the black square  in Figure 9 (see the five global minima therein). 

\vskip \baselineskip
\noindent
Finally, for $K/J > 0.25$, all ordered phases disappear, remaining only phase $\bf P$. Therefore, the last topology is that shown in Figure 9. In the next section we summarize the results of this study. 

\section{Conclusions}

We have studied, in a linear chain of $N$ spins ($N \to \infty$), the spin-$3/2$ Blume-Capel model with competing long-and short-range interactions, and anisotropy $D$ ($D>0$). Conveniently, the long-range interactions were represented by  mean-field  ferromagnetic couplings ($J>0$), and the short-range interactions were represented by  antiferromagnetic couplings ($K>0$) between  nearest-neighbor spins. We obtained the phase diagrams in the $D/J-k_{B}T/J$ plane, for different values of $K/J$, so as to explore how the topology of the   well-known spin-$3/2$ Blume-Capel model with mean-field ferromagnetic couplings is modified as $K/J$ increases. As a first step for understanding the results at finite temperatures, we obtained the phase diagram in the $K/J-D/J$ plane for $T=0$. Four magnetic orderings are present, namely, two ferromagnetic phases $\bf F_{1}$ and $\bf F_{2}$, with $|S_{i}|=3/2$ and $|S_{i}|=1/2$, respectively, and two antiferromagnetic phases $\bf AF_{1}$ and $\bf AF_{2}$, with  $|S_{i}|=3/2$ and $|S_{i}|=1/2$, respectively. For finite temperatures and without competition $K/J=0$, the phase diagram in the $D/J-k_{B}T/J$ plane  contains an upper second-order frontier dividing the  ferromagetic region and the paramagnetic region. The ferromagnetic region is composed by the two ordered phases $\bf F_{1}$ and $\bf F_{2}$  separated by a fisrt-order line ending at an ordered critical point, which is bellow the second-order line. We started the competition by increasing the value of $K/J$. So, the topology of the phase diagram is changed by the modification of the original second-order frontier  and the first-order one. Thus, the frontier dividing the ferromagnetic and the paramagnetic region remains of second-order approximately for $0 < K/J < 0.1758 $. Then, for $0.1785 < K/J < 0.2250$, this line is divided by three sections two of them are of first-order, an intermediate second-order section limited by two tricritical points. The tricritical points approach themselves as $K/J$ increases, so, for $K/J > 0.2250$ this frontier is only of first-order.  On the other hand, $K/J < 0.1875$, the first-order frontier that divides phases $\bf F_{1}$ and $\bf F_{2}$ does not suffer any change, however, for greater values of $K/J$ a branch line of first-order emerges from it, ending at an ordered critical point too. This branch line  encloses partially a new  phase $\bf F_{3}$, whose region is ordered in such a way that the mean magnetization per spin is mostly  $|m|=1$.  This must be because for each spin $i$ with $S_{i}=3/2$($S_{i}=-3/2$), there is another  spin $j$ with $S_{j}=-1/2$ ($S_{j}=1/2$), such that the total spin sums one (minus one). This configuration minimizes the free energy density in that region of the phase diagram. Thus, The branch line grows as $K/J$ increases, and both ending points of the lines dividing phases $\bf F_{1}$, $\bf F_{2}$ and $\bf F_{3}$ approach the upper first-order frontier.  Finally, for $K/J > 0.24585$, the phase $\bf F_{3}$ is completely enclosed when the ending points touch the upper frontier, so the ordered region is now divided in three separated zones corresponding to  $\bf F_{1}$, $\bf F_{2}$ and $\bf F_{3}$. Therefore, this last topology contains three points of coexistence. The lower point, at which the branch line begins, meets phases $\bf F_{1}$, $\bf F_{2}$ and $\bf F_{3}$, whereas the upper points, one on the left and the other on the right,  meet phases $\bf F_{1}$, $\bf F_{3}$ and $\bf P$, and $\bf F_{3}$, $\bf F_{2}$ and $\bf P$, respectively. For $K/J > 0.25$, all ordering disappears, and only phase $\bf P$ is present in the phase diagram.

\vskip \baselineskip

{\large\bf Acknowledgments}

Financial support from CNPq (Brazilian agency) is acknowledged. 

\vskip \baselineskip

\vspace{0.5cm}
\begin{figure}[htbp]
\centering
\includegraphics[width=6.0cm,height=5.0cm]{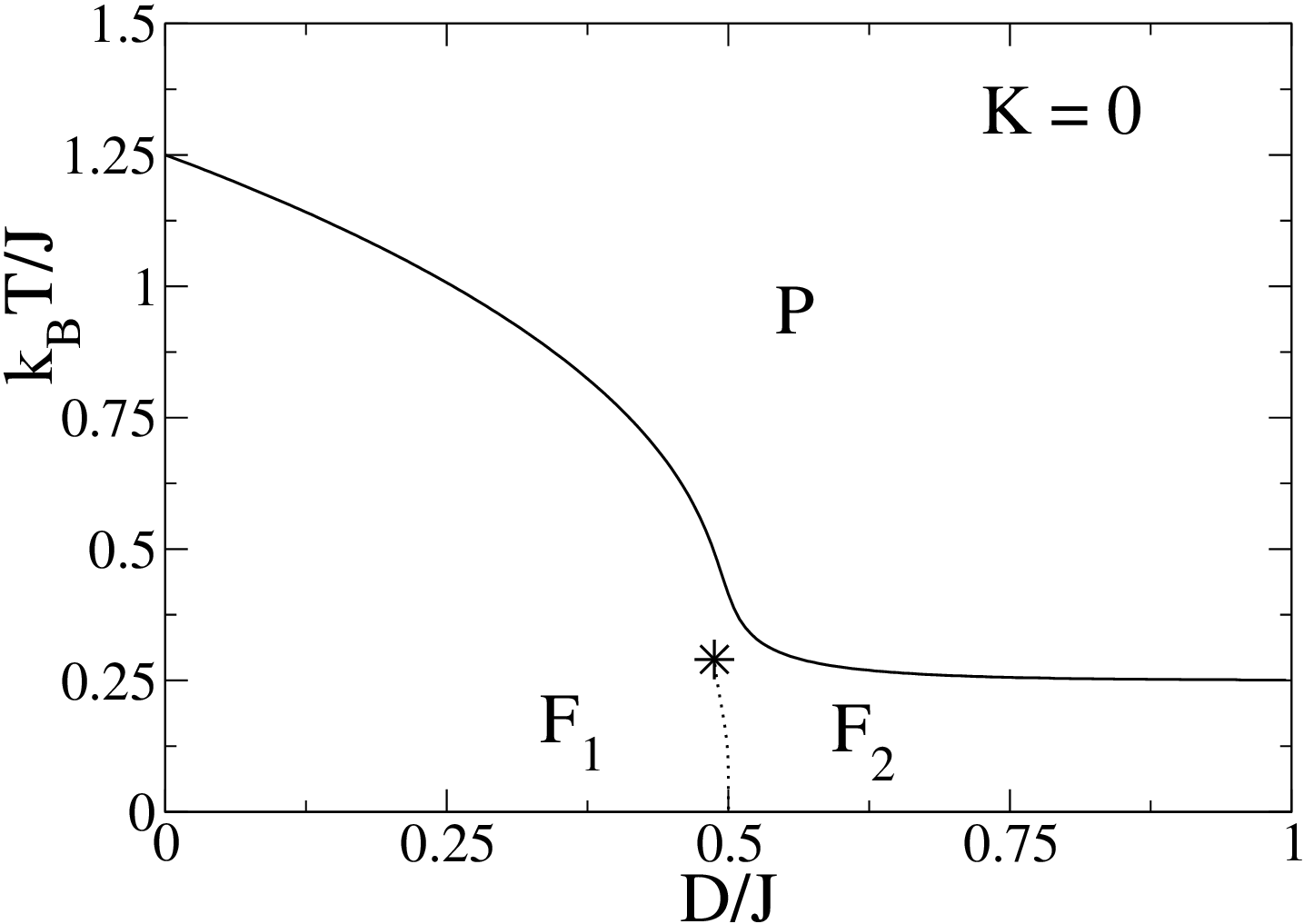}
\caption{Phase diagram of the spin-$3/2$ Blume Capel Model with  mean-field ferromagnetic interactions. } 
\label{figura1}
\end{figure}

\vspace{3.0cm}
\begin{figure}[htbp]
\centering
\includegraphics[width=5.5cm,height=4.5cm]{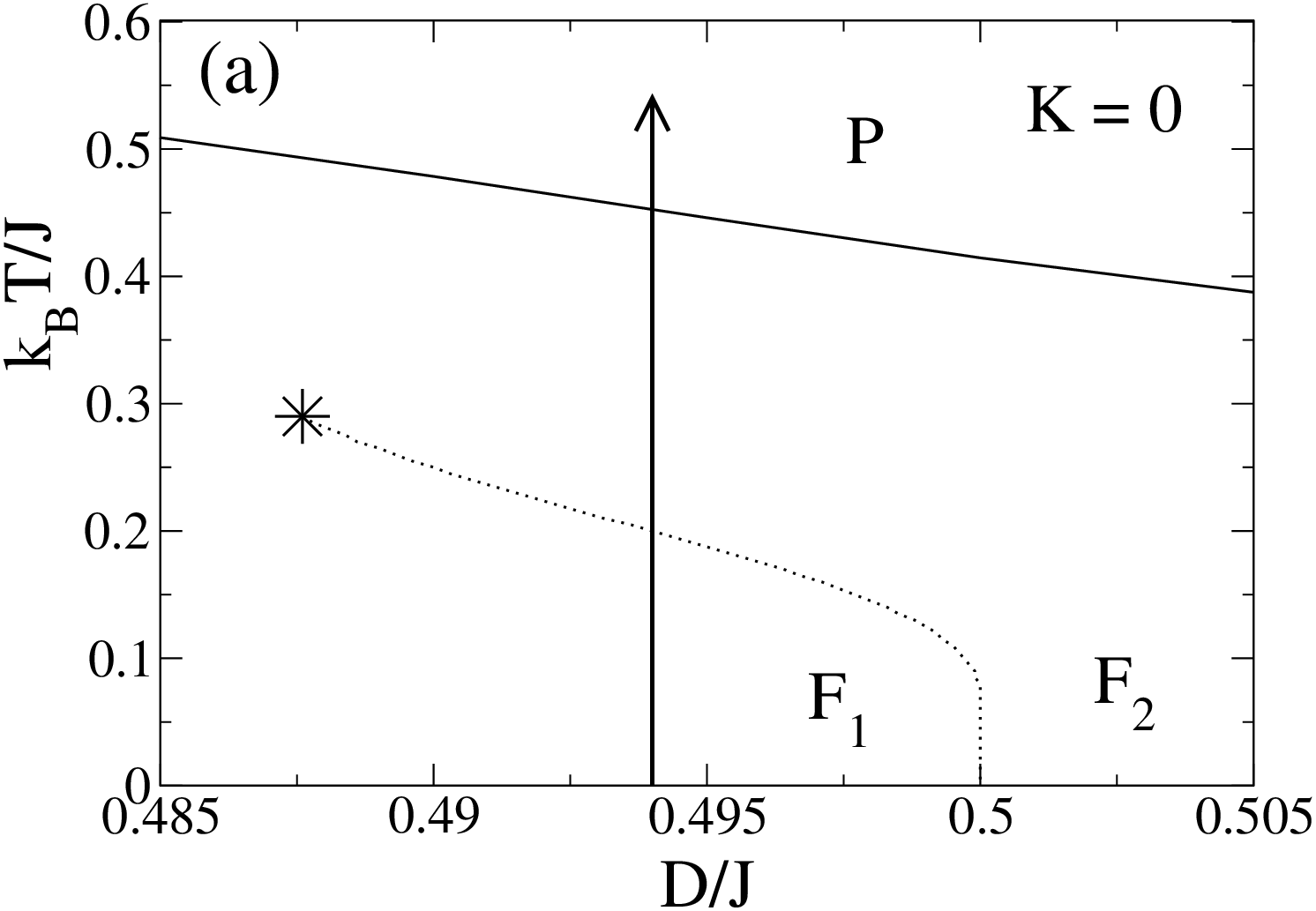}
\hspace{0.5cm}
\includegraphics[width=5.5cm,height=4.5cm]{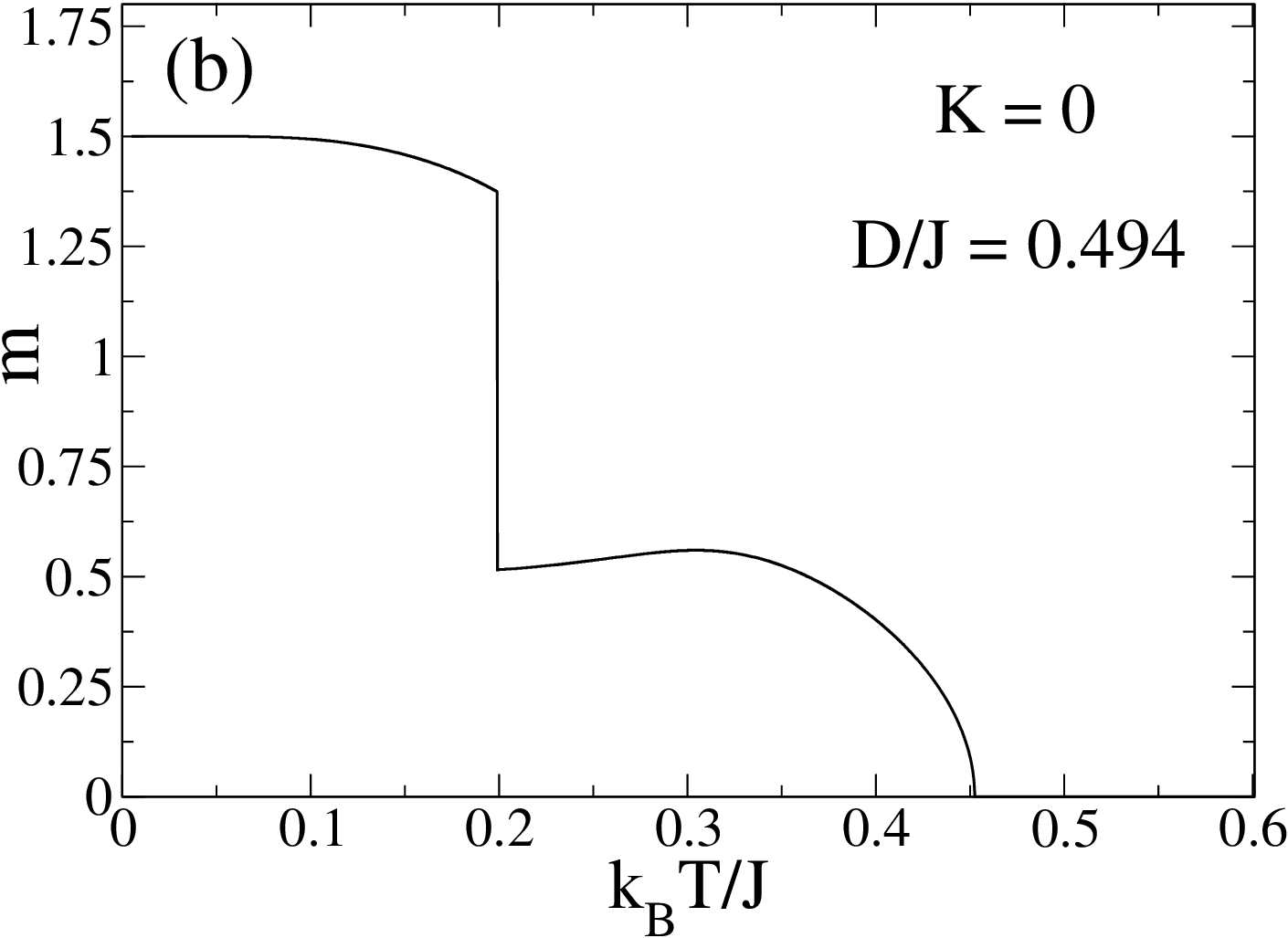}
\caption{(a) Portion of the phase diagram of spin-$3/2$ Blume Capel Model showing the coexistent line that separates phases $\bf F_{1}$ and $\bf F_{2}$. 
The arrow is a guide to the eyes for marking  the line for which the magnetization is plotted in (b); (b) The magnetization versus the temperature for $D/J = 0.494$. The jump discontinuity is a signal of a first-order phase transition between phases $\bf F_{1}$ and $\bf F_{2}$. The magnetization increases after crossing the line of coexistence, until reaching a maximum value. Then it decreases continuously downto zero, signaling a second-order phase transition between phases $\bf F_{2}$ and $\bf P$.  } 
\label{figura2}
\end{figure}

\vskip \baselineskip
\noindent
%%%%%%%%%%%%%%%%%%%%%%%%%%%%
\begin{figure}[htp]
\begin{center}
\includegraphics[height=6.0 cm]{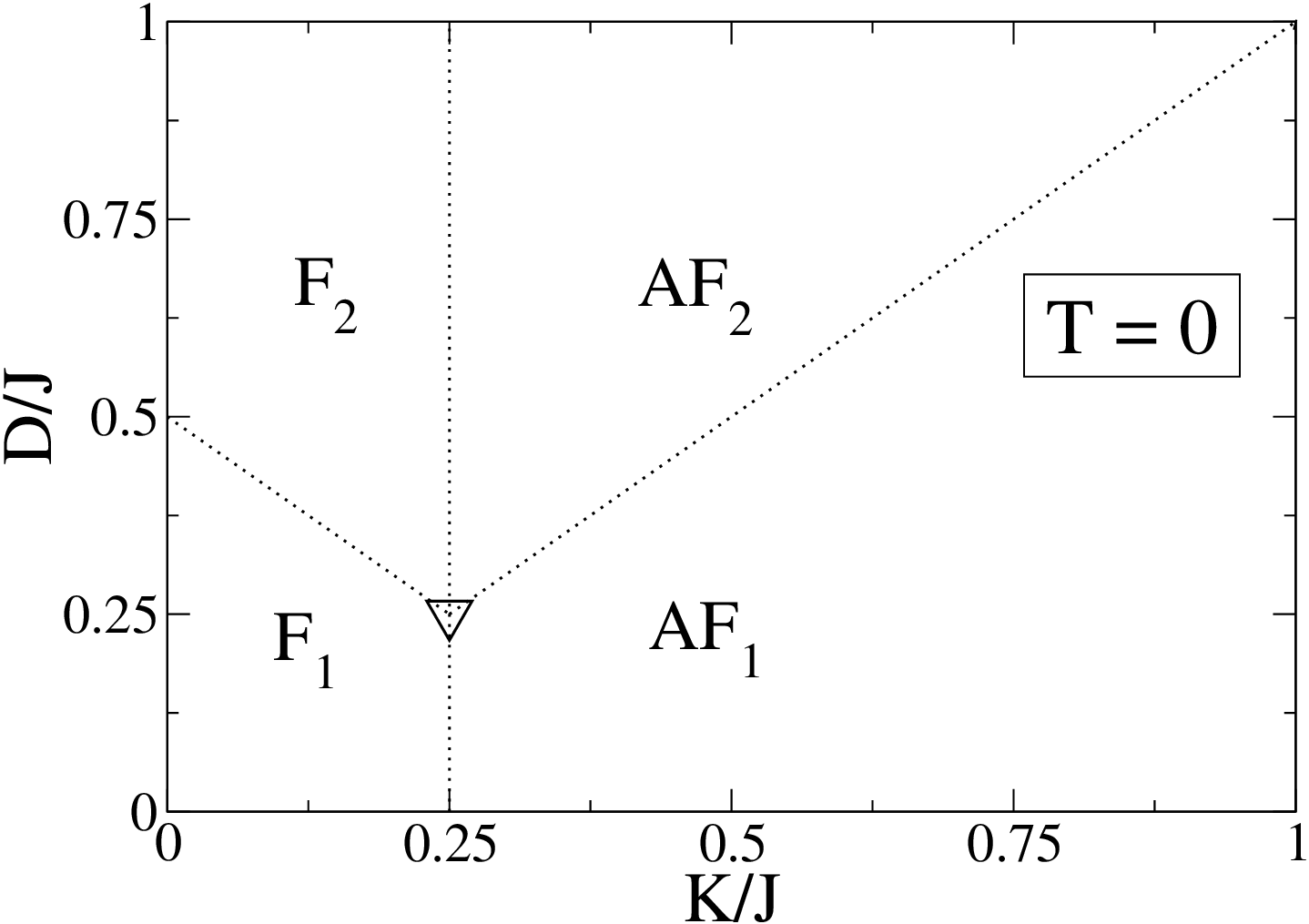}
\end{center}
\vspace{-1cm}
\caption{\footnotesize Phase Diagram of the ground state of the Model. The three lines are  first-order frontiers meeting  at a triple point represented by the empty diamond. Phases $\bf F_{1}$ and $ \bf F_{2}$, are ferromagnetic phases corresponding to magnetizations per spin, $|m|=3/2$ and $|m|=1/2$, respectively. In the antiferromagnetic phase $\bf AF_{1}$, spins align in a regular pattern with neighboring spins pointing in opposite directions,  with $|S_{i}|=3/2$, for $i=1,2,...,N$,  whereas in the antiferromagnetic phase $\bf AF_{2}$ the absolute value of each spin is $|S_{i}|=1/2$, for $i=1,2,...,N$.  }
\label{figura3} 
\end{figure}
%%%%%%%%%%%%%%%%%%%%%%%%%%%%

\vspace{3.0 cm}
\begin{figure}[htbp]
\centering
\includegraphics[width=6.5cm,height=5.5cm]{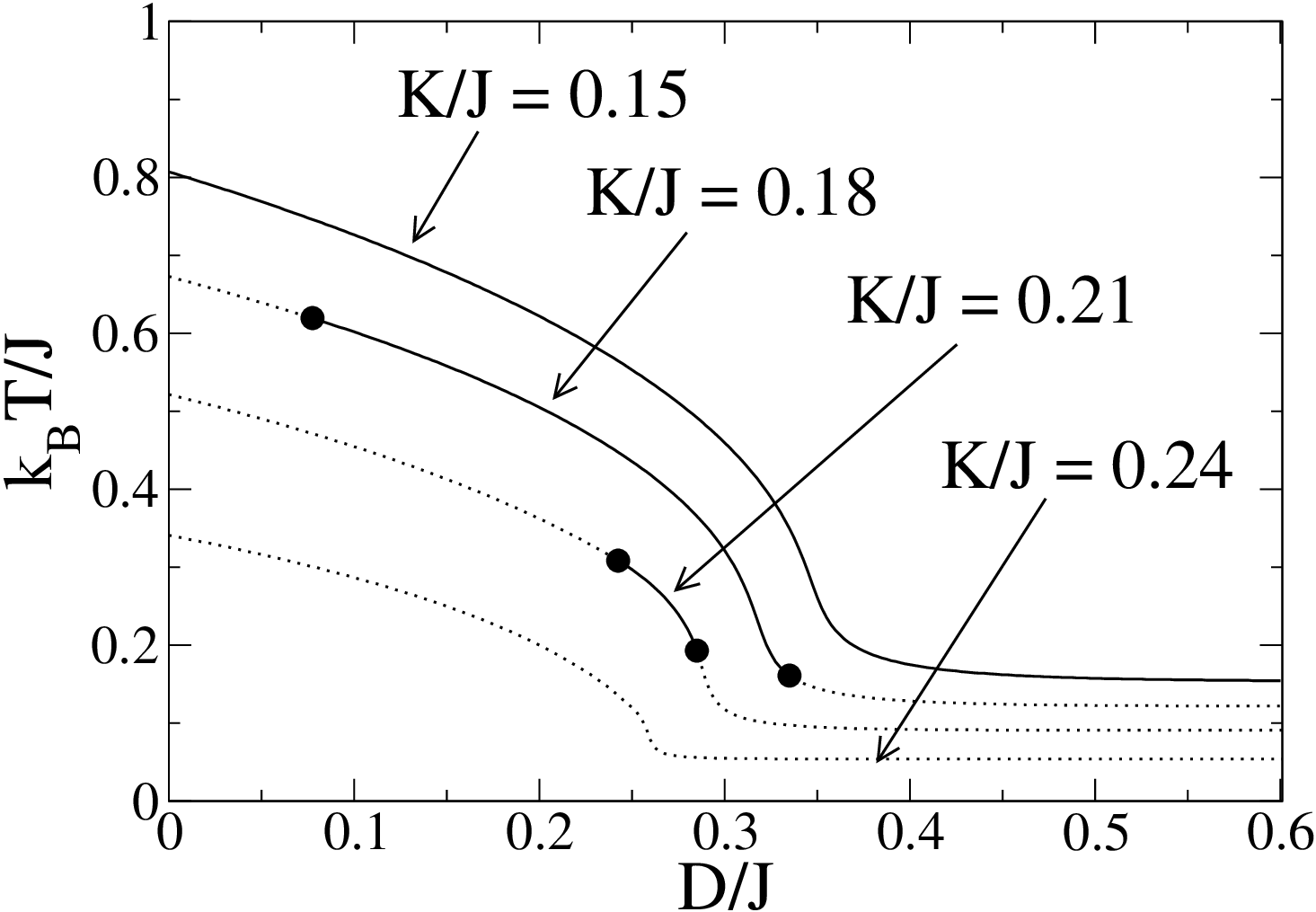}
\caption{Evolution of the original second-order frontier of the  spin-$3/2$ Blume-Capel model when $K$ increases. Note that this frontier is only of second order for lower values of $K/J$ ($K/J < 0.1756$). However, for greater values of $K/J$, the frontier is now divided into three sections. The  second-order section is limited by two tricritical points, and the two other ones are of first-order in the extremes (see the dotted  portions). The second-order section is reduced when $K/J$ increases. This happens until $K/J$ reaches  certain critical value ($K/J \simeq 0.225$). Then, the frontier is only of first-order.} 
\label{figura4}
\end{figure}

\vspace{3.0 cm}
\begin{figure}[htbp]
\centering
\includegraphics[width=6.5cm,height=4.5cm]{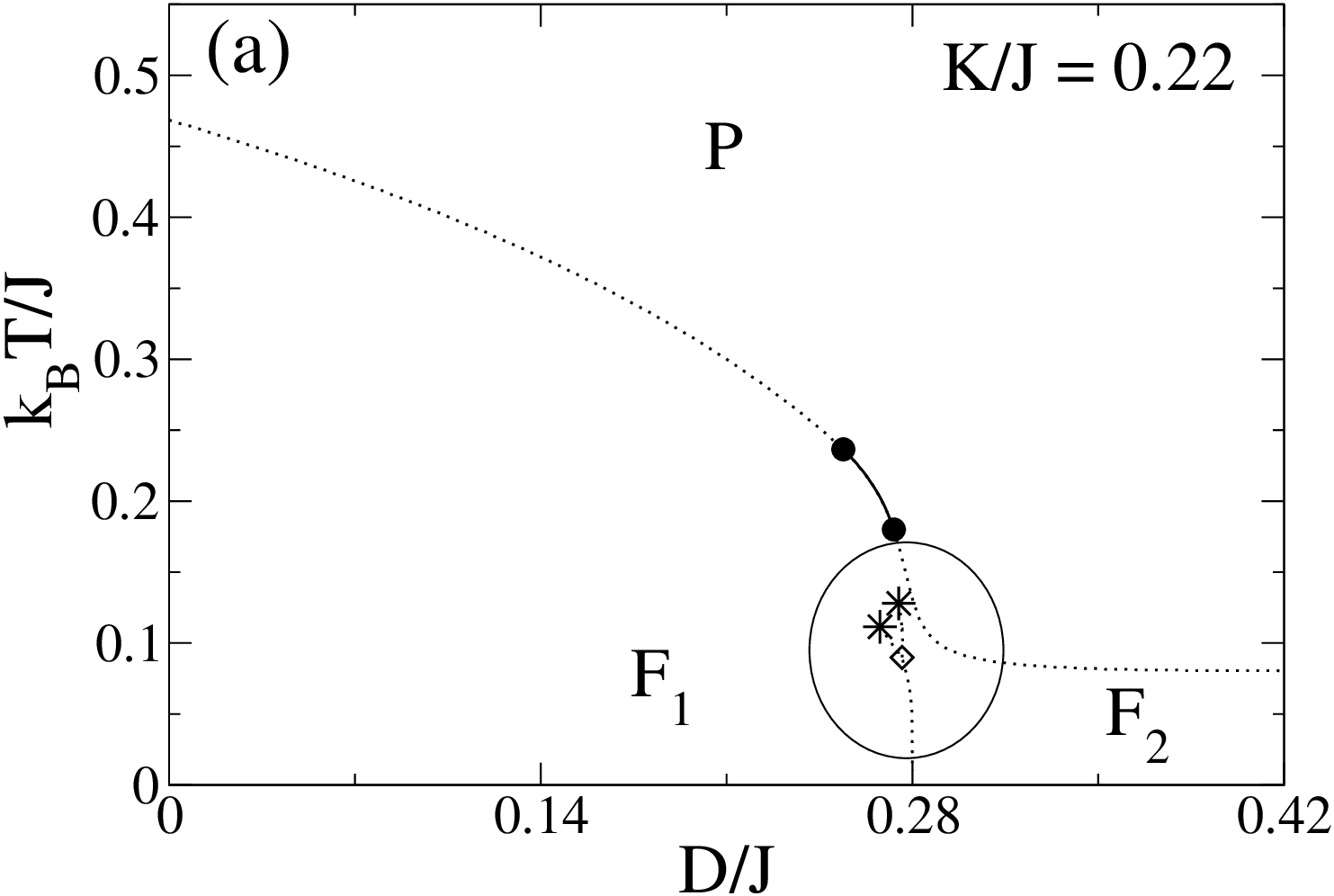}
\hspace{0.5cm}
\includegraphics[width=6.5cm,height=4.5cm]{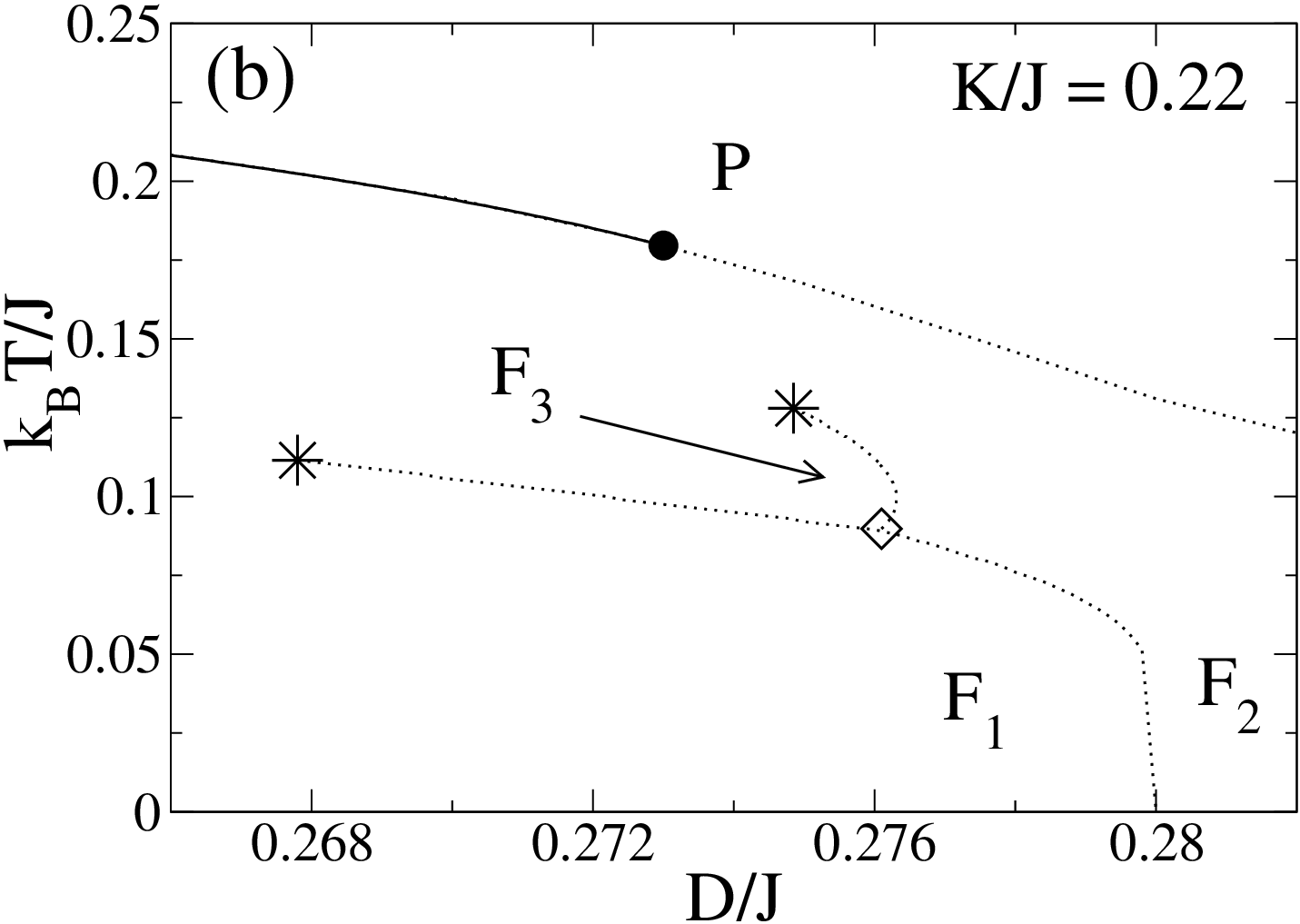}
\caption{(a) Phase diagram of the model for $K/J=0.22$, showing the appearance of a branch line of first-order emerging from the coexistence line that separates phases $\bf F_{1}$ and $\bf F_{2}$ (the circle is a guide to the eye to highlight the region of interest). (b) A portion of the phase diagram shown in (a), in which we observe more clearly the branch line enclosing a third ordered phase $\bf F_{3}$.     } 
\label{figura5}
\end{figure}

\vspace{0.5cm}
\begin{figure}[htbp]
\centering
\includegraphics[width=8.0 cm,height=7.0cm]{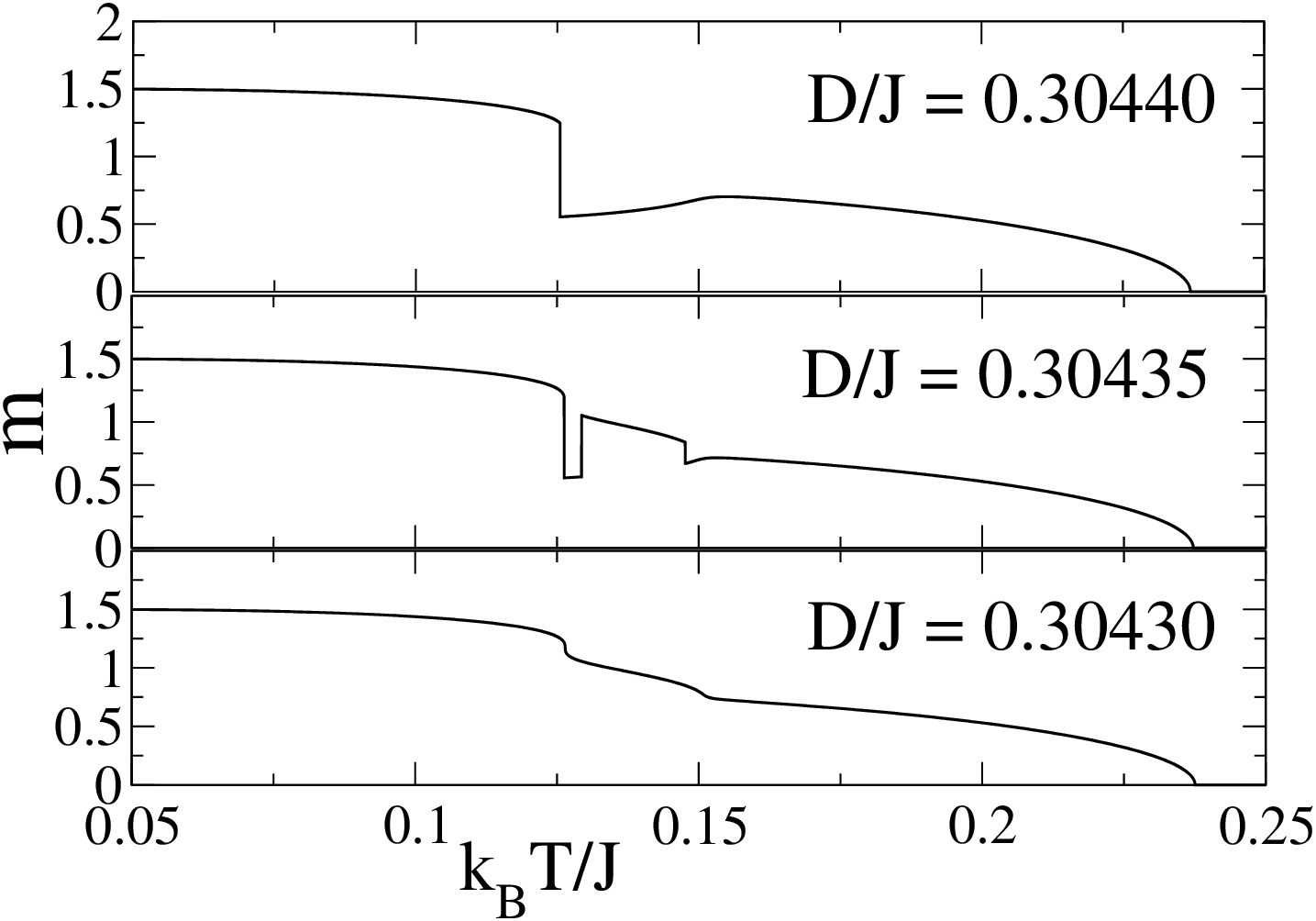}
\caption{The magnetization curve for three close values of $D/J$, for $K/J=0.19$. For $D/J=0.30430$, the whole magentization curve is continuous because it has not crossed any first-order line.    For $D/J=0.30435$, the magnetization curve suffers three jump discontinuities, signaling the presence of the branch line (see it in Figure 5, for $K/J=0.22$). For $D/J=0.30440$, the magnetization suffers only one jump discontinuity, which shows  that it has crossed only one line of coexsistence. Accordingly, for $K/J=0.19$, the range of the branch line of first-order is too short ($\Delta D/J < 0.0001$). Thus, its onset in the phase diagram must be for a value of $K/J$ just less than  $K/J=0.19$. } 
\label{figura6}
\end{figure}

\vspace{4.0 cm}
\begin{figure}[htp]
\centering
\includegraphics[width=5.5cm,height=4.5cm]{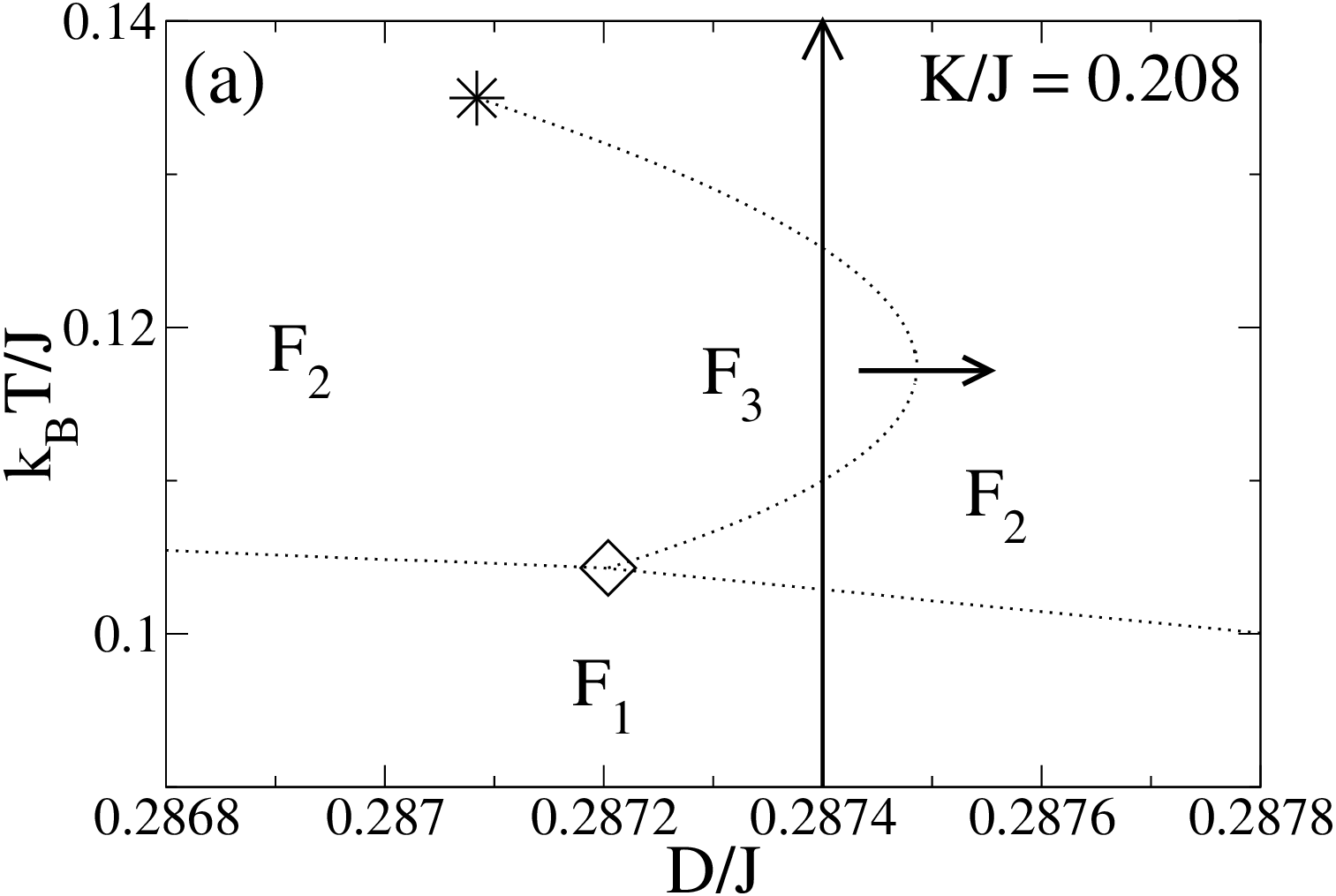}
\hspace{0.25 cm}
\vspace{0.7cm}
\includegraphics[width=5.5cm,height=4.5cm]{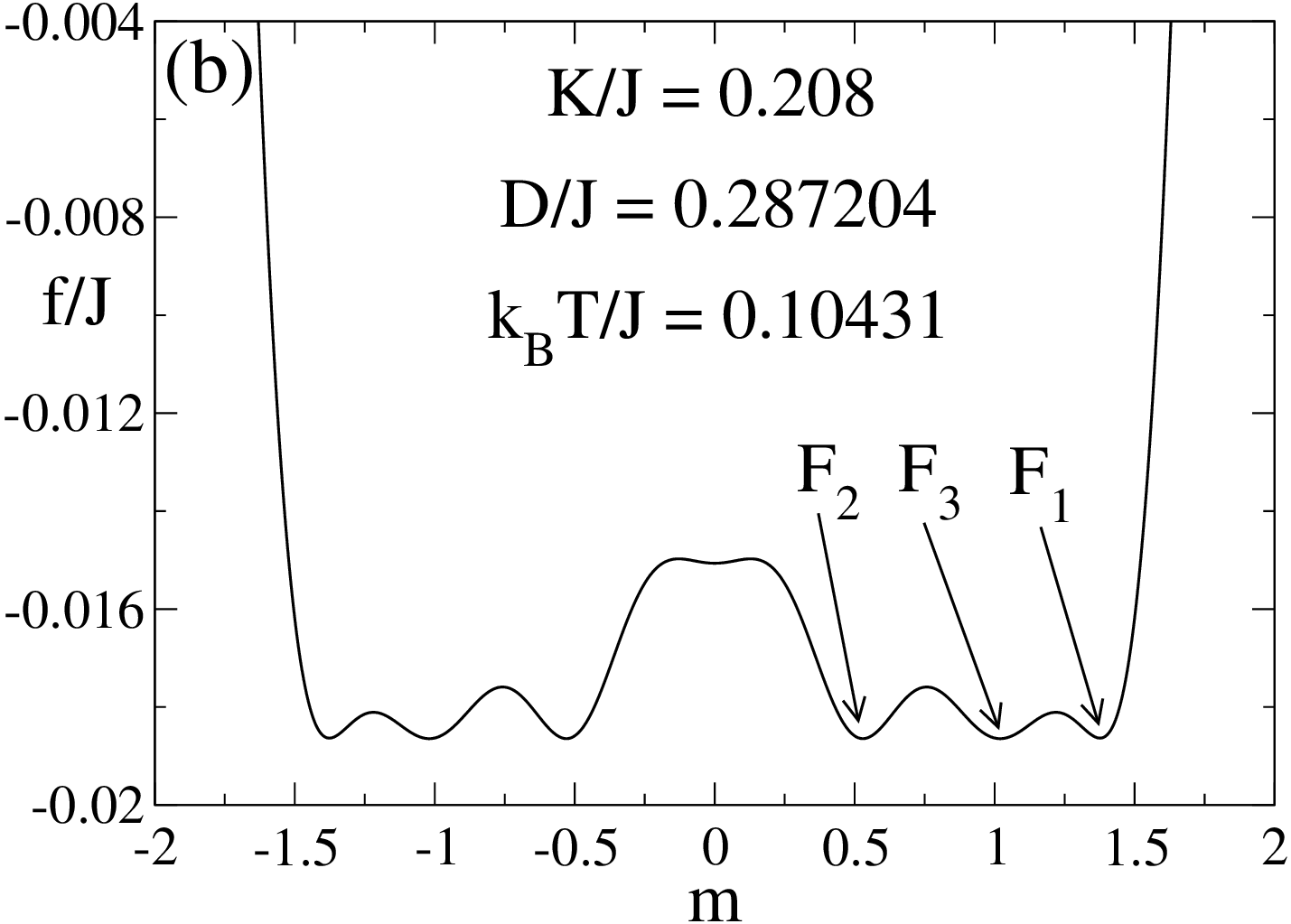}
\hspace{5.0cm}
\includegraphics[width=5.5cm,height=4.5cm]{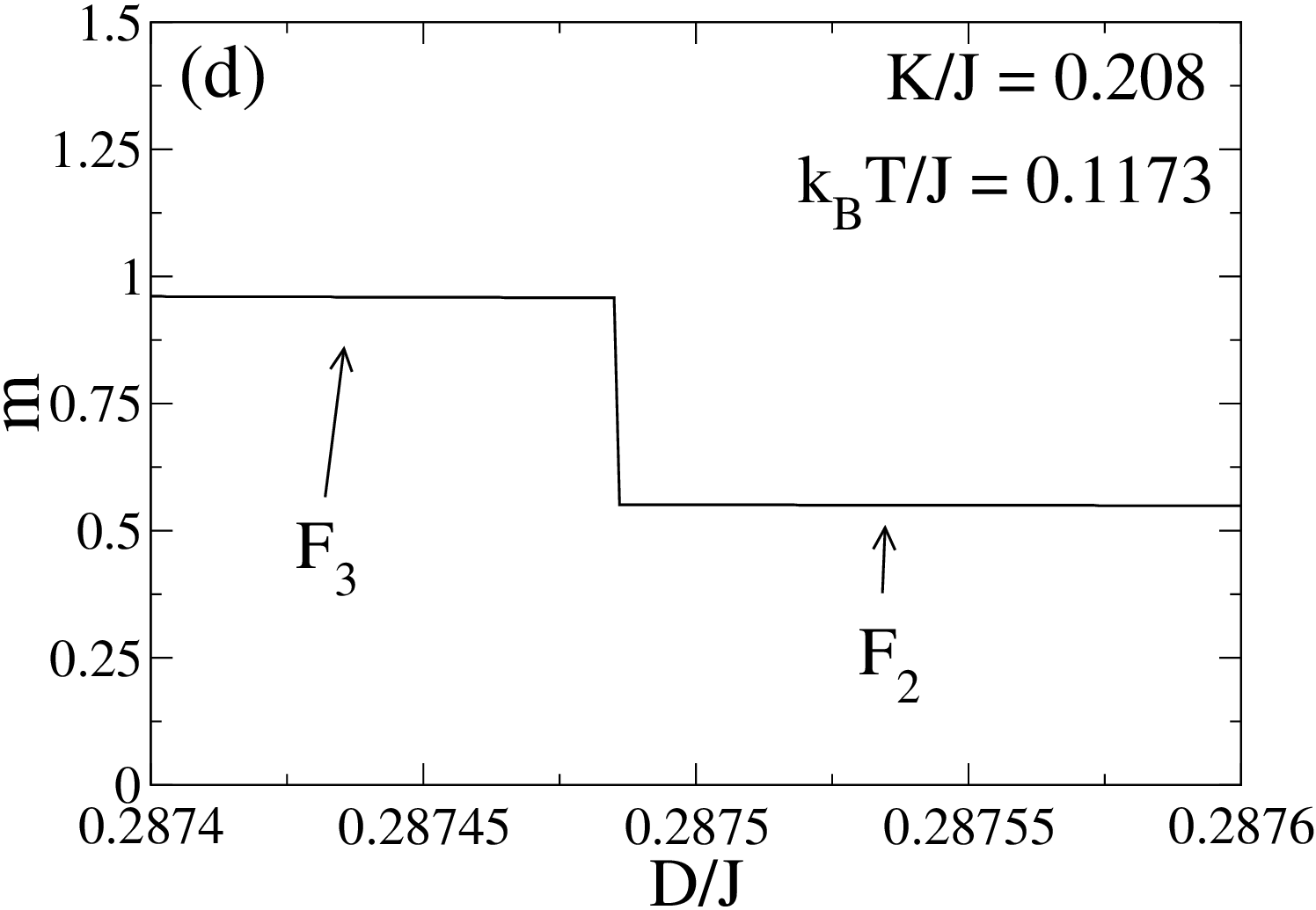}
\hspace{0.25 cm}
\includegraphics[width=5.5cm,height=4.5cm]{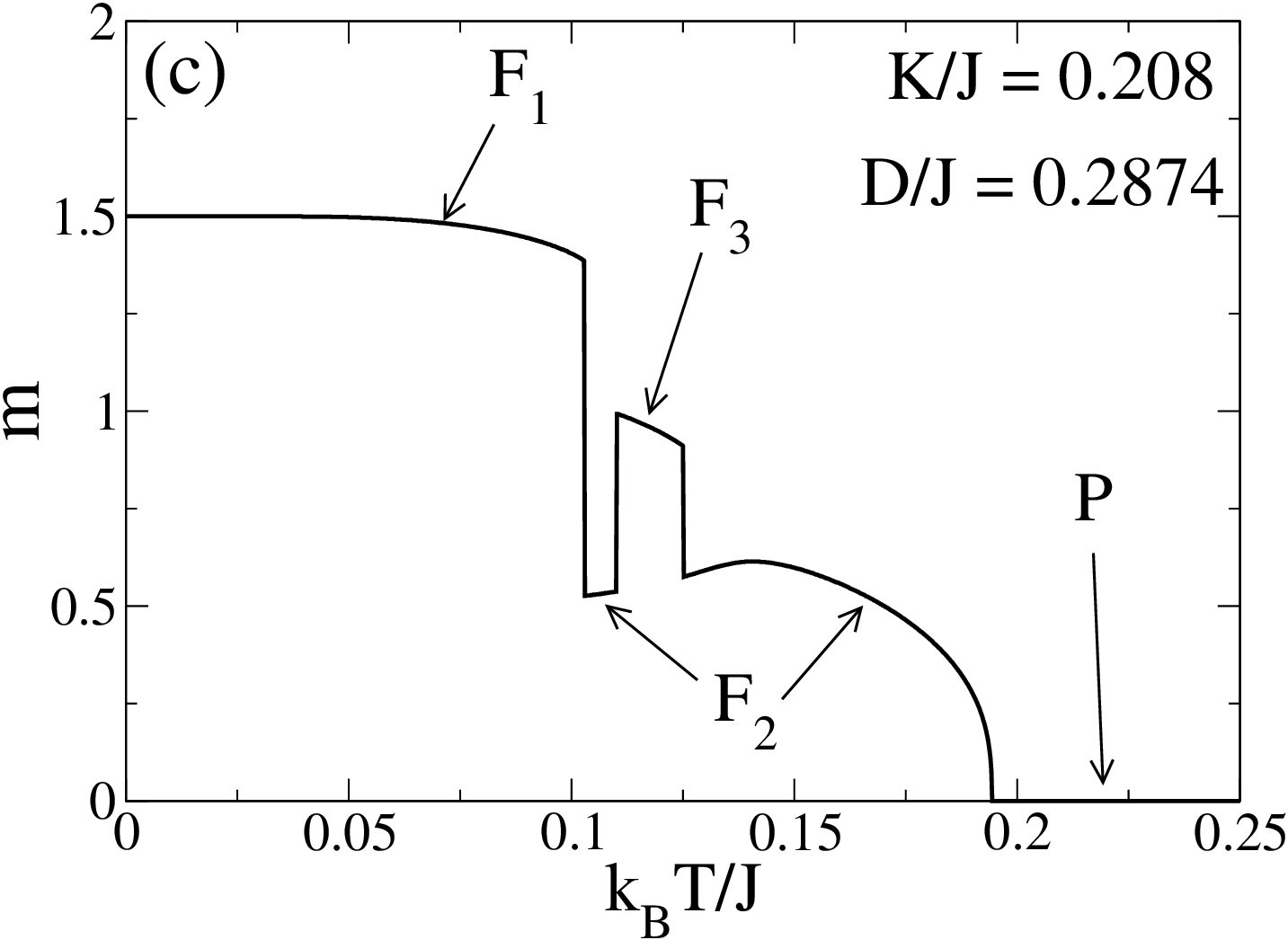}
\caption{(a) A portion of the phase diagram, for $K/J=0.208$, showing the reentrant behavior of the branch line of first-order enclosing the phase $\bf F_{3}$. The arrows are guide to the eye for marking where the magnetization is plotted in (c) and (d). In  (b), The free energy density versus the magnetization per spin, at the coexistent point represented by the empty diamond shown in (a). We can observe six symmetric minima at the same level. This shows that phases $\bf F_{1}$, $\bf F_{2}$ and $\bf F_{3}$ coexist at this point. In (c), the magnetization curve versus the temperature plotted for the points marked by the vertical arrow shown in (a). In (d), the magnetization curve versus the temperature plotted for the points signalized by the horizontal arrow shown in (a). } 
\label{figura7}
\end{figure}

\vspace{2.5cm}
\begin{figure}[htbp]
\centering
\includegraphics[width=7.0cm,height=5.5cm]{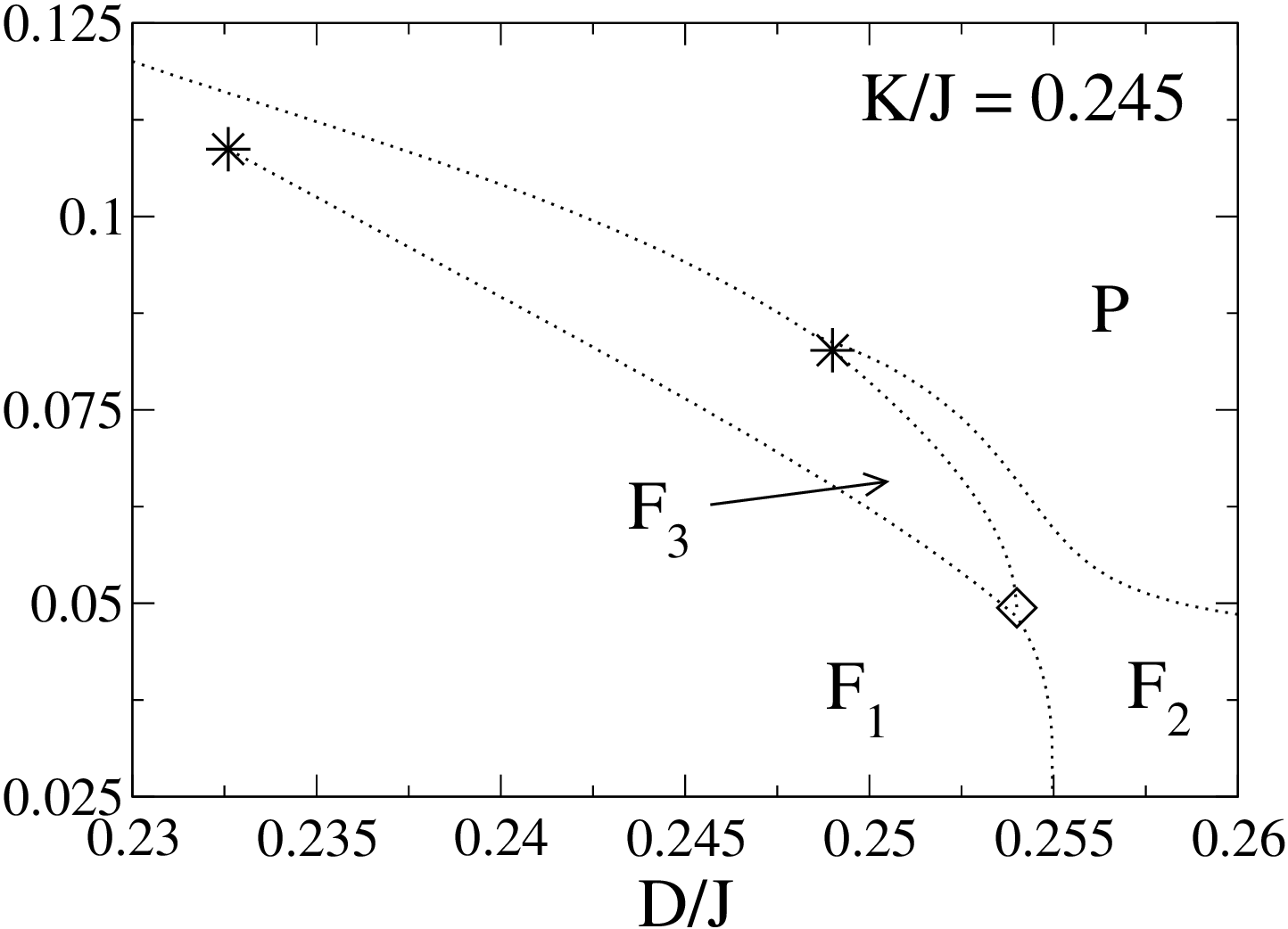}
\caption{Phase diagram of the model for $K/J=0.245$. The ending points represented by the asterisks are very close to the first-order frontier dividing the ferromagnetic and the paramagnetic phases. It can also be observed that the reentrancy of the branch line has disappeared.   } 
\label{figura8}
\end{figure}

\vspace{2.5cm}
\begin{figure}[htbp]
\centering
\includegraphics[width=7.5cm,height=5.5cm]{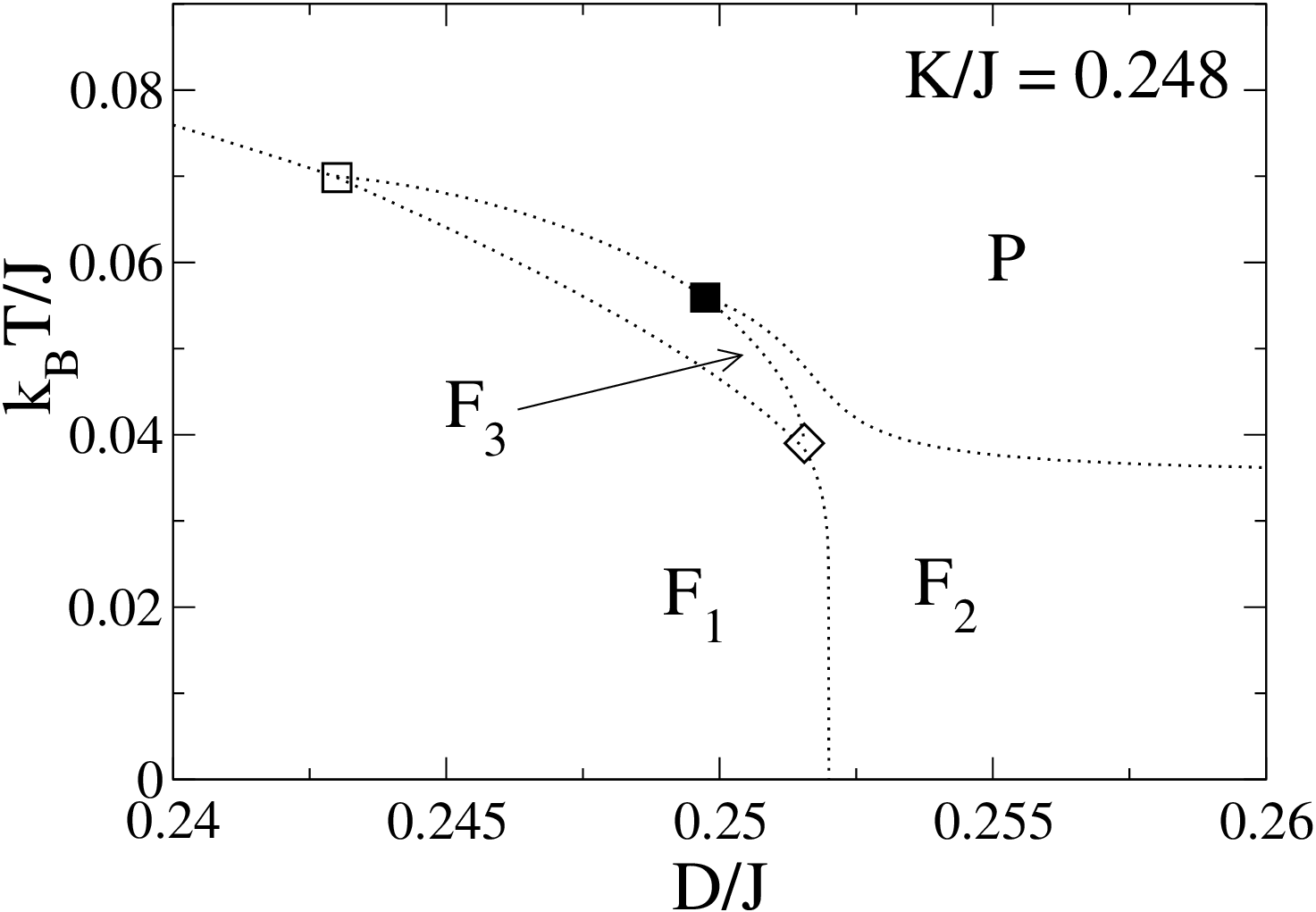}
\caption{Phase diagram of the present model, for $K/J=0.248$. The ending points of the first-order frontiers dividing phases $\bf F_{1}$, $\bf F_{2}$ and $\bf F_{3}$, are now at the first-order frontier that separates the ferromagnetic and the paramagnetic regions of the phase diagram. These ending points are represented by the empty square and the black square. So,  phases $\bf F_{1}$, $\bf F_{3}$ and $\bf P$ coexist at the empty square, whereas phases $\bf F_{3}$, $\bf F_{2}$ and $\bf P$ coexist at the black square.   } 
\label{figura9}
\end{figure}

\vspace{3.0 cm}
\begin{figure}[htbp]
\centering
\includegraphics[width=5.5cm,height=4.5cm]{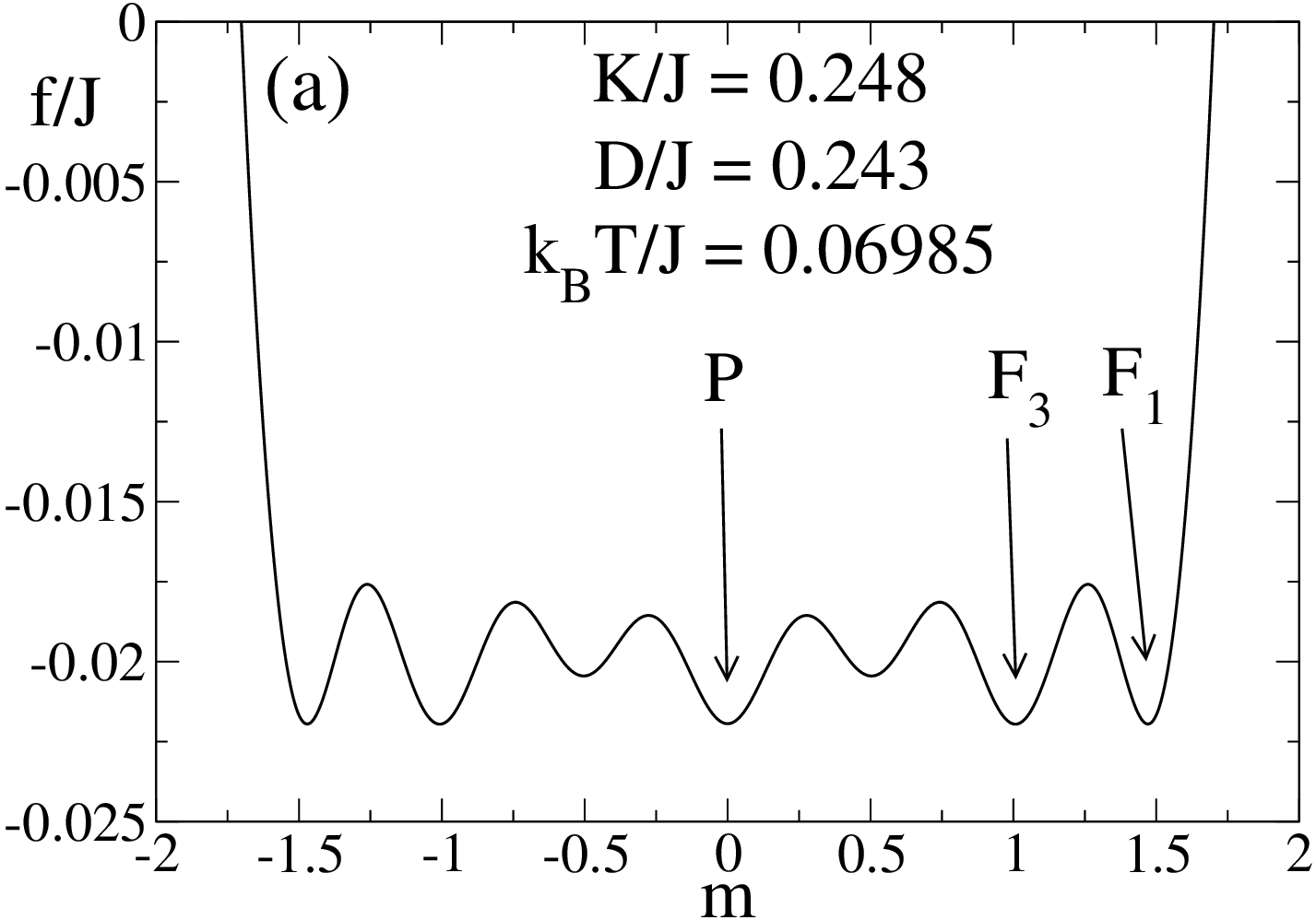}
\hspace{0.5cm}
\includegraphics[width=5.5cm,height=4.5cm]{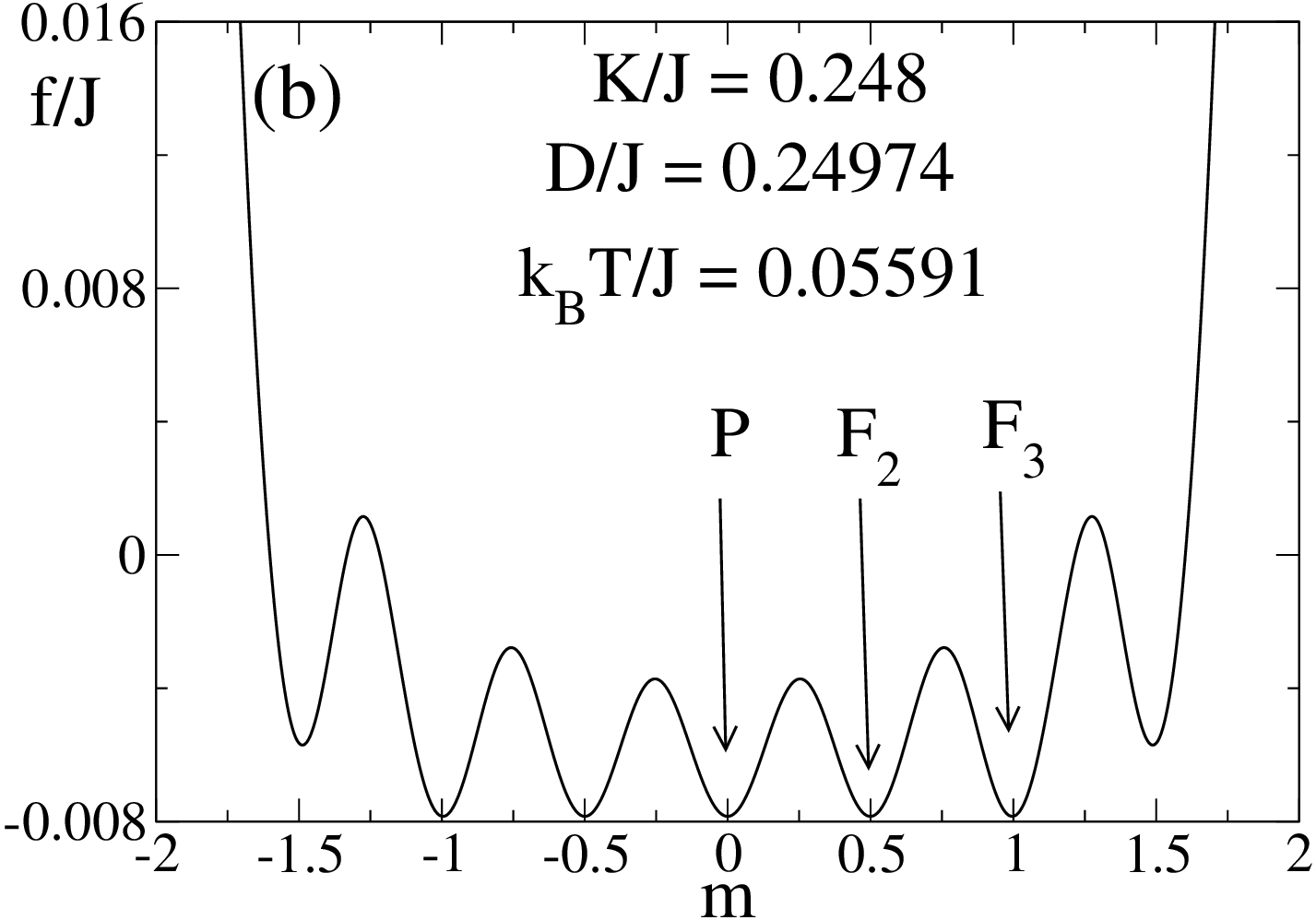}
\caption{(a) The free energy density at the point of coexistence represented by empty square in Figure 9; (b) The free energy density at the point of coexistence represented by the black square in Figure 9. The values of $m$ at the global minima compose the coexisting spin phases at equilibrium, while the other minima correspond to metastable states.   } 
\label{figura10}
\end{figure}

\end{document}